\documentclass[aps,twocolumn,groupedaddress,superscriptaddress,showpacs,pra,floatfix]{revtex4}

\usepackage{amsmath}
\usepackage{graphicx}
\usepackage{dcolumn}
\usepackage{bm}
\usepackage{epsfig}
\usepackage{dsfont,psfrag}
\usepackage{color}
\usepackage{ifthen}

\newcommand{\ket}[1]{{| #1\rangle}}

\newcommand{\E}{\mathrm{e}}

\begin{document}
  \title{Robustness of {spin-chain} state-transfer schemes}
\author{Joachim Stolze}
\email{joachim.stolze@tu-dortmund.de}
\affiliation{Technische Universit\"at Dortmund, Fakult\"at Physik, D-44221 Dortmund, Germany}
\author{Gonzalo A. \'{A}lvarez}
\affiliation{Department of Chemical Physics, Weizmann Institute of Science,  76100 Rehovot, Israel}
\email{gonzalo.a.alvarez@weizmann.ac.il}
\author{Omar Osenda}
\affiliation{Facultad de Matem\'{a}tica, Astronom\'{i}a y F\'{i}sica, Universidad Nacional de C\'{o}rdoba, 5000 C\'{o}rdoba, Argentina}
\email{osenda@famaf.unc.edu.ar}
\author{Analia Zwick}
\affiliation{Department of Chemical Physics, Weizmann Institute of Science,  76100 Rehovot, Israel}
\email{analia.zwick@weizman.ac.il}
\date{\today}
\begin{abstract}
This is a shortened and slightly edited version of
  a chapter \cite{my64} in the collection {\em Quantum State Transfer and Network
    Engineering}, edited by G.M. Nikolopoulos and I. Jex \cite{NJ13}, where
 we review our own research about the
robustness of spin-chain state-transfer schemes along with other approaches to the
topic. Since our own research is documented elsewhere to a large
extent we here restrict ourselves to a review of other approaches
which might be useful to other researchers in the field.
\end{abstract}

\maketitle

\section{Introduction}
\label{sec:1}

\subsection{Spin chains}
\label{subsec:1.1}

The ground state of a one-dimensional ferromagnetic spin-1/2 chain is
the all-up state $
  \ket{\uparrow}_1 \otimes \ket{\uparrow}_2 \otimes ... \otimes
  \ket{\uparrow}_N
= \ket{\uparrow \uparrow ... \uparrow} .
$
Here, the states $\ket{\uparrow}_i$ and $\ket{\downarrow}_i$ are the
eigenstates of the Pauli spin operator $\sigma_i^z$ acting at site $i$
of an $N$-site chain, with eigenvalues +1 and -1,
respectively.  A very simple excited state is then created by
flipping a single spin; $\sigma_j^- \ket{\uparrow \uparrow
  ... \uparrow}$, where $\sigma_j^- = \frac12 (\sigma_j^x -i
\sigma_j^y)$. This state breaks translational invariance, which may be
restored, however, by superposition:
\begin{equation}
  \label{eq:2}
  \ket{\psi_k} := \frac 1{\sqrt{N}}  \sum_{j=1}^N \E^{ikj} \sigma_j^-
\ket{\uparrow \uparrow ... \uparrow} .
\end{equation}
Here $k=\frac{2 \pi \nu}{N} \; (\nu=0,...,N-1)$ is a dimensionless
wave number, and we have temporarily assumed periodic boundary
conditions. The state (\ref{eq:2}) is called a
(single-) spin-wave state in magnetism \cite{Mat81}. Spin waves are often found to be low lying
excited energy eigenstates of spin chain models, with the
energy-momentum relation (dispersion relation) $\omega(k)$ depending
on the detailed nature of the model Hamiltonian. 

Information-carrying signals must be localized in space and time.
The Fourier relation (\ref{eq:2}) may be inverted to
represent a localized single spin-flip state $
  \sigma_j^- \ket{\uparrow \uparrow ... \uparrow} 
$
as a superposition of spin-wave states $\ket{\psi_k}$. 
This is the most sharply localized state available in a spin
chain.

In quantum information science the single spin-1/2
system is called a quantum bit (qubit), its states $\ket{\uparrow}$
and $\ket{\downarrow}$ are mapped to the basis states $\ket{0}$ and
$\ket{1}$ of the qubit, respectively, and the all-up ferromagnetic state is mapped
to the computational basis state $\ket{\mathbf 0} = \ket{00...0}$ of the $N$-qubit chain. 
The single spin-flip state  then is interpreted as one of the
$N$ computational basis states with a single non-zero qubit at site
$j$: $
   \sigma_j^- \ket{\uparrow \uparrow ... \uparrow}
   =: \ket{\mathbf j} .
$ As stated above that state may be expressed as a
superposition of spin wave states (\ref{eq:2}) each of which
propagates in space-time with amplitude $\E^{i(kj-\omega t)}$, that is,
with a phase velocity $
  v_{\phi}=\frac{\omega(k)}k
$ which in general varies with $k$ if the dispersion relation
$\omega(k)$ is nonlinear. Dispersion then takes its toll by broadening
and flattening the pulse in the course of time.


In the remainder of this
introduction we will sketch some of the main lines of attack on the
problem of quantum information transfer via quantum spin chains. All
those proposals work well under ideal ``design''
conditions. Robustness of some of the schemes under less than ideal
conditions is discussed subsequently. 

\subsection{Quantum information transfer in ideal quantum spin chains}
\label{subsec:1.2}
\begin{widetext}
To set the stage we define the following general nearest-neighbor
coupled spin-1/2 chain Hamiltonian:
\begin{equation}
  \label{eq:6}
  H=\frac 12 \sum_{i=1}^N J_i \left[
(1+\gamma)\sigma_i^x \sigma_{i+1}^x +
(1-\gamma)\sigma_i^y \sigma_{i+1}^y +
\Delta \sigma_i^z \sigma_{i+1}^z
\right] 
+\sum_{i=1}^N h_i  \sigma_i^z   .
\end{equation}
\end{widetext}
We identify site 1 with the fictitious site $N+1$, that is, $\sigma_{N+1}^{\alpha} \equiv \sigma_{1}^{\alpha} \;(\alpha=x,y,z)$.
 Then $H$ describes a ring for  $J_N \ne 0$ and an open chain of $N$ spins for  $J_N=0$.
 If $J_i$ (for $i \ne N$) and $h_i$ do not depend
on $i$ we call the system homogeneous, otherwise it is
inhomogeneous. The symmetry of the spin-spin interaction is controlled by the
anisotropy parameters $\Delta$ and $\gamma$. 
For $\gamma=0$ and $\Delta=1$ 
the model is the original Heisenberg model, also known as XXX model,
because all spin components experience the same coupling to their
nearest neighbors. 
For $\gamma=0$ and $\Delta \ne 1$ 
the model is called XXZ model, the important special case $\Delta=0$
being known  as XX model \footnote{Some authors also use the name XY
  model; we would like to reserve the term XY model for the
  anisotropic case, $\Delta=0, \gamma \ne 0$.}. 
$\gamma=0$ implies conservation of the total $z$ spin component, a
case of obvious importance in quantum information transfer, since the
number of ones in the state of the $N$-qubit system as expressed in
the computational basis is then a constant of the
motion. If  $\gamma \ne 0$ and $\Delta \ne0, 1$ the model is referred
to as the general XYZ chain. 
Finally, for $\gamma \ne 0$ and $\Delta =0$ 
the model is known as the XY chain, a special case being $\gamma =
\pm1$ where we have an Ising chain in a transverse magnetic field (TI
chain). 
The $\Delta=0$ case of (\ref{eq:6}) may be mapped
        \cite{LSM61,Kat62}          to a
model of noninteracting spinless lattice fermions with
nearest-neighbor hopping by means of a Jordan-Wigner
transformation
\cite{JW28}, with the number of fermions being conserved for the XX
case, $\gamma=0$.  Hence the XX chain has been quite popular as a
model for quantum information transfer due to its simple dynamics.

The earliest example \cite{Bos02}  for quantum information
transfer in a spin chain, however, employed a ferromagnetic Heisenberg
chain in its ground state with homogeneous couplings to whose ends a
``sender'' and a ``receiver'' spin can be coupled by the
experimenters, Alice and Bob, after Alice has prepared the sender spin
in the single-qubit state she wishes to transmit. As the state of the
combined system evolves, information is transferred to the other end of
the chain, where Bob at some suitable time decouples the receiver spin
from the rest of the system and measures or further processes it. The
fidelity of this process is less than perfect, but higher than the
maximal value of 2/3 \cite{HHH99} for classical transmission of a
quantum state, for $N \leq 80$. Osborne and Linden \cite{OL04} pointed
out the deleterious effect of dispersion on single-qubit quantum state
transfer and proposed to encode the quantum information to be
transmitted  not in the state of a single spin, but in a spin-wave
packet constructed so as to involve only the approximately linear part
of the dispersion relation $\omega(k)$ of the spin-wave excitations in
a homogeneous Heisenberg model ring. Burgarth et al. \cite{BB04,BGB04}
suggested to improve the Heisenberg chain transfer protocol by using
two or more chains in parallel and performing measurements in order to
increase the fidelity of transfer. That ``multirail''
protocol\index{multirail protocol} is
treated in \cite{BG14}. 
Besides the presence of
dispersion effects the homogeneous Heisenberg chain has another
disadvantage which limits its usefulness and which was pointed out by,
among others, Subrahmanyam \cite{Sub04}: two flipped spins in the
ground state of a ferromagnetic Heisenberg chain interact with each
other, leading to distortions of a two spin-flip state. More
technically speaking, this means that while single spin-wave states
are energy eigenstates, two spin-wave states are not, making their
treatment more complicated and ultimately leading to the intricacies
of the Bethe Ansatz \cite{Bet31}.

Interactions between elementary excitations can be avoided by going
from the Heisenberg chain to the XX chain by dropping the $z$ part of
the nearest-neighbor interaction, but dispersion cannot be avoided as
long as homogeneous chains ($J_i \equiv J$ in (\ref{eq:6})) are
considered. This precludes {\em perfect state 
transfer\index{perfect state transfer}} (to be discussed below)
for all but very short chains \cite{CDEL04}. However, recent work on
{\em pretty good state transfer\index{pretty good state transfer}} \cite{GKSx12} shows that arbitrarily
patient observers will obtain fidelities arbitrarily close to 1 in
homogeneous XX chains of length $N=P-1, 2P-1$ with $P$ a prime, or
$N=2^M-1$. The waiting time depends on the required deviation from
unity and seems to grow exponentially in the chain length.

Haselgrove \cite{Has05} discussed an interesting scheme involving
time-dependent manipulation of the nearest-neighbor couplings between
both sender and receiver qubits and the rest of the
chain.  Another attempt at improving the transfer
fidelity of homogeneous quantum chains is the application of a sequence of
two-qubit gates (switchable interactions) between the end of the
spin-chain ``wire'' and the receiving qubit \cite{BGB07}; some results
on the stability of this scheme against disorder were reported in
\cite{Bur07}.  These
schemes involving time-dependent couplings or external fields are at
the border of quantum optimal control\index{quantum optimal control} theory, an extremely rapidly
developing field of research which, however, is not within the scope
of the present chapter.

Zenchuk \cite{Zen12} suggested to consider not perfect state transfer
but ``complete information transfer'': The state of the sender system
$S$ (a subsystem of the communication system under study) is encoded
in the initial reduced density operator $\rho_S(0)$ of $S$. The
quantum time evolution of a fairly arbitrary chain system then
performs a linear mapping of $\rho_S(0)$ to $\rho_R(t)$, the reduced
density operator of the receiver subsystem $R$. If the Hilbert space
of $R$ is at least as large as that of $S$ the linear mapping
$\rho_S(0) \mapsto \rho_R(t)$ can be inverted for almost all times $t$
and the original information may be reconstructed.

The comprehensive review by Bose \cite{Bos07}
covers the development of spin-chain quantum information transfer until
the end of 2006. The special class of homogeneously coupled spin
chains is discussed in \cite{BBS+14}.

Since quantum information transfer in strictly homogeneous chains
suffers from dispersion effects as explained above, two main
strategies have been developed, both based on the natural dynamics of
inhomogeneous qubit chains. One approach uses ``fully
engineered\index{fully-engineered spin chain}''
chains, where ${\cal O} (N)$ coupling constants or local fields have to
be assigned specific values in order to achieve perfect state transfer
(PST). The other approach employs
``boundary-controlled\index{boundary-controlled spin chain}'' chains in
which only ${\cal O} (1)$ coupling constants connecting sender and receiver
qubits to the transmitting ``wire'' have to be adjusted in order to
achieve optimized state transfer (OST). Both strategies will be
explained below.

\subsubsection{Perfect state transfer\index{perfect state transfer} in fully engineered chains}
\label{subsubsec:1.2.1}

Perfect state transfer in engineered chains is based on the 
observation that a quantum system generates periodic dynamics if its
energy spectrum displays only
commensurate energy differences. 
Examples are the harmonic oscillator with energies $E_n=\hbar \omega
(n+ 1/2) \; (n \ge 0)$ or the infinite square well with 
$E_n=n^2 E_1  \; (n \ge 1)$. A popular exercise in elementary quantum
mechanics shows that arbitrary wave functions develop periodically (up
to a global phase) under a harmonic force: $
  \psi(x,t)=-\psi\left(x,t+\frac{2 \pi}{\omega}\right)  .
$ More interesting in the present context, but less often discussed in
quantum mechanics courses is the relation $
  \psi\left(x,t+\frac{ \pi}{\omega}\right) = -i \psi(-x,t)  ,
$ meaning that in one half period the state of the oscillator develops
into a perfect spatial mirror\index{mirror} image of the original state. This
mirroring property of the quantum oscillator 
rests on the commensurate energy spectrum and on
the alternating parities of successive energy eigenstates. If these
two properties can be carried over to a one-dimensional array of
quantum mechanical objects that array can be used for perfect quantum
state transfer.

A charged spin-$J$ particle in a magnetic field in $z$ direction shows
a finite equidistant energy spectrum of $2J+1$ levels, the energy
eigenstates being the eigenstates of $J_z$. Transitions between these
states are caused by the transverse spin components $J_x$ and
$J_y$. In a 1979 paper Cook and Shore \cite{CS79} employed the analogy
between a spin-$J$ system and a $(2J+1)$-level atom to derive a model
for stepwise laser excitation, obtaining periodic solutions which
permitted complete population inversion. Nikolopoulos et
al. \cite{NPL04b} turned population inversion to spatial inversion by
suggesting an array of quantum dots\index{quantum dot array} able to accomodate an electron
each, with tunneling between neighboring quantum dots adjusted so as to
mimic the dynamics of the spin-$J$ system. The same line of thought
was followed by Christandl et al. \cite{CDEL04} who found PST in
an inhomogeneous open $N$-spin XX chain with couplings
$J_i=\sqrt{i(N-i)}$ and equidistant energy levels in the single
spin-flip sector. Albanese et al. \cite{ACDE04} generalized the
concept to other state-mirroring systems using earlier results
\cite{AJNx98} on the construction of finite quantum systems with
periodic dynamics. Shi et al. \cite{SLSS04} discussed other spin chains
involving external fields and displaying a commensurate energy
spectrum. 

Yung and Bose \cite{YB04} and  Karbach and Stolze \cite{my44}
suggested a systematic approach to the construction of PST chains with
a desired energy spectrum by solving a special type of Jacobian
inverse eigenvalue problem\index{inverse eigenvalue problem}  \cite{Gla04,CG05}.  Given a set of $N$ real numbers $E_1 < E_2 < ... <E_N$, it
is always possible to construct a unique {\em persymmetric\index{persymmetric} Jacobi
  matrix}, that is, a real symmetric tridiagonal matrix with diagonal
entries $a_1, ...,a_N$ and strictly positive super- / subdiagonal
entries $b_1, ... ,b_{N-1}$, with the additional symmetry conditions
$a_i=a_{N+1-i}$ and $b_i=b_{N-i}$ which has the prescribed numbers
$E_i$ as eigenvalues. Note that the number of given eigenvalues equals
the total number of independent matrix elements $a_i$ and $b_i$. In an
XX chain the $b_i$ are related to the nearest-neighbor couplings $J_i$
and the $a_i$ are related to the local fields $h_i$ in
(\ref{eq:6}). The persymmetry of the matrix corresponds to the spatial
symmetry of the spin chain and makes sure that successive eigenvalues
correspond to eigenvectors of opposite parities. The eigenvalue
spectrum can be chosen freely, as long as the energy differences are
commensurate, to ensure PST. That freedom of choice may be used to
provide the system with other desirable properties besides PST. In
\cite{my44} it was shown, for example, that it is possible to perform
PST in chains with nearly homogeneous couplings, deviating from
$J_i=$const only on the few percent scale. If the diagonal matrix
elements $a_i$ (the local fields $h_i$ in the XX chain) vanish, the
eigenvalues are symmetrically distributed about zero, that is, $E_i=
\pm |E_i|$ and the number of unknowns in the inverse eigenvalue
problem is greatly reduced. For that special case, a simple algorithm
was recently proposed by Wang et al. \cite{WSR11}; for the more
general case, many interrelated algorithms are known
\cite{Gla04,CG05}. 


A simple spin-1/2 XX chain\index{XX model} with PST thus is given by 
\begin{equation}
  \label{eq:9}
  H=\frac 12 \sum_{i=1}^{N-1} J_i (\sigma_i^x \sigma_{i+1}^x +
  \sigma_i^y \sigma_{i+1}^y) \quad (J_i=J_{N-i})   ,
\end{equation}
with the $J_i$ chosen appropriately.
Let us pause briefly to point out
some important facts about this system. By the Jordan-Wigner
transformation\index{Jordan-Wigner transformation} \cite{JW28} (\ref{eq:9}) is equivalent to the
Hamiltonian of noninteracting lattice fermions with hopping elements
$J_i$. The number of fermions corresponds to the number of down spins
and the ferromagnetic all-up state corresponds to the fermion vacuum. The Jacobi
matrix discussed above in the context of the inverse eigenvalue
problem is the Hamiltonian restricted to the subspace of a single
fermion. All single-fermion states are transferred to their spatial
mirror images at the same instant of time, and so are suitable many-fermion
states, since the Jordan-Wigner fermions do not interact. This means
that in contrast to many other quantum information transfer schemes,
PST in the system  (\ref{eq:9}) is not restricted to single-qubit
states, as already noted in \cite{ACDE04}. 

A generalization of the above PST systems was suggested by Kostak et
al. \cite{KNJ07} who discuss Hamiltonians which generate
permutations between the sites of a network and can thus be employed
for single-qubit perfect state transfer. A generalization of the PST
concept in the mathematical sense is ``almost PST'' as defined
recently \cite{VZ12} by replacing the periodic functions in the time
evolution of the quantum system with almost periodic functions. A useful
review on various aspects of PST was written by Kay \cite{Kay10}.

\subsubsection{Optimized state transfer\index{optimized state
    transfer} in boundary-controlled chains\index{boundary-controlled spin chain}}
\label{subsubsec:1.2.2}

The second strategy employing  spin chains,
boundary-controlled chains, is restricted to the transfer of
single-qubit states by construction. In this strategy the sender qubit
$A$ and the receiver qubit $B$ are only weakly coupled to the large
system used to transfer the quantum information, the ``quantum data
bus\index{quantum data bus}'', for example, a homogeneous spin chain.


 An early example for that concept was given in \cite{PS05}, for a 
 ring of coupled harmonic oscillators. Within the single-excitation subspace
the coupled oscillator chain is equivalent to a spin-1/2 XX chain. 
A very clear picture of what is going on in the
weak-coupling\index{weak boundary coupling} scenario
was given in the short paper by W\'ojcik et al. \cite{WLK+05}. There a
homogeneous $N$-site XX chain with $J_i \equiv 1$ is coupled to two
end spins with coupling strength $\alpha<1$. The eigenvalues and
eigenvectors responsible for single-excitation transfer are determined
analytically and it is found that for very small values of $\alpha$
only two or three (when $N$ is odd) closely-spaced energy levels in
the center of the the spectrum are important. Oscillations between
these states determine the quantum information transfer, and the
transfer time increases as $\alpha$ decreases, while the deviation of
the fidelity from unity scales as $\alpha^2N$. The transfer time is
${\cal O}(\alpha^{-2})$ for even $N$ and ${\cal O}(N^{\frac 12}
\alpha^{-1})$ for odd $N$. Furthermore it is observed that after half
the transfer time  the two end spins become entangled for even $N$.

Another system proposed \cite{LSC+05} to be used as a data bus is
a spin ladder. Due to the excitation gap above the ladder's ground
state, perturbation theory can be used to eliminate the ladder to
lowest order and to replace it by an effective coupling between qubits
$A$ and $B$. The ground-state entanglement between the end spins in some spin
chains was suggested \cite{VBR07} to be used for teleportation or
state transfer even at finite temperature $T$, as long as $T$ is
smaller than the smallest excitation gap which depends on the chain
length. Under suitable conditions, end spins $A$ and $B$ weakly
coupled to an intermediate chain can then be approximated by an
effective two-spin model. Similar approximations are also considered
in more recent spin bus scenarios \cite{OSFx12}. The same general strategy, of
separating the sender-receiver Hilbert space of the spins $A$ and $B$
from the rest of the system, was followed in \cite{GKM+08}. 
Since the nature of the transfer medium connecting $A$
and $B$ is to a large extent irrelevant, 
that medium may even be a spin chain with random
couplings, according to \cite{YJG+11}. The separation of end spins
from the remainder of the system may also be achieved by applying
strong local fields to the end spins \cite{CLMS09}. Thereby, two
states of opposite parities and both strongly localized at the
boundary spins are created, which are spectrally separated from the
remainder of the Hilbert space, and which may be used for near-perfect
state transfer \cite{my61}. This last example shows very clearly a
drawback common to many of the boundary-controlled scenarios. The
smaller the coupling between sender / receiver and data bus, the
better is the quality of transfer. At the same time, unfortunately,
the energy differences driving the dynamics become smaller and smaller
and the transfer slows down, increasing the danger of the quantum
information being destroyed by fluctuating interactions with the
environment before it is transferred completely.

The boundary-controlled scenarios discussed up to now may be termed
``weak-coupling'' scenarios. We now briefly discuss the
``optimal-coupling'' scenario discovered in a numerical study by Zwick
and Osenda \cite{ZO11} and at the same time developed analytically by
Banchi et al. \cite{BACx10,BACx11}.
 As in \cite{WLK+05}, a homogeneous $N$-site XX chain with $J_i
\equiv 1$ is coupled to two end spins with coupling strength
$\alpha<1$, however, $\alpha$  is optimized in a different way. For
$\alpha=1$, i.e. a completely homogeneous XX chain with $N+2$ sites, the energy
eigenvalues in the single-excitation subspace are proportional to
$\cos k$, with $k=\frac{n\pi}{N+3}, (n=1, ... , N+2)$, while the
corresponding eigenvector amplitudes at site $i$ of the chain are 
$u_k(i) = \sqrt{\frac{2}{N+3}} \sin ki$. Note that in the vicinity of
$k=\frac{\pi}2$ the energies are approximately linear in $k$, so that
packets\index{wave packet} of spin waves from that region are approximately free from
dispersion. For $\alpha \ne 1$ the situation changes, but it is still
possible to treat the eigenvalue problem analytically. It turns out
\cite{BACx11} 
that the initial state with the sender spin down (and  all others up)
corresponds to a wave packet with approximately Lorentzian probability
distribution in $k$, with center and width depending on
$\alpha$. Since the energy spectrum also depends on $\alpha$ it is
possible to jointly optimize the width of the wave packet and the
linearity of the energy spectrum in order to achieve near-perfect
state transfer. It turns out that the optimal value of $\alpha$ scales
as $N^{-\frac 16}$. Further optimization is possible \cite{ABC+12} if
not only the first and last, but also the second and second to last
bonds may be adjusted.

\section{Fully-engineered vs. boundary-controlled chains}
\label{sec:2}


The topic of the present book chapter is the transfer of quantum
information\index{quantum information transfer} solely by the natural dynamics of a spin chain\index{spin chain} with fixed
couplings, as explained in section \ref{subsec:1.2}. Although the
perfect-transfer schemes of section \ref{subsubsec:1.2.1} are able
to transfer multi-qubit states, we shall restrict our attention to the
most frequently discussed case, 
single-qubit state transfer. Since
nothing in this world is perfect, including computer hardware and
software, both classical and quantum, information transport in spin
chains is vulnerable to two main sources of irregularity: external
dynamic randomness, that is, fluctuating fields caused by the
environment, and internal static randomness, caused by inaccurate
implementation of the theoretical design of the chain. Here we shall
exclusively deal with static randomness and with the robustness of both the
perfect state transfer\index{perfect state transfer} chains from section  \ref{subsubsec:1.2.1}
and the optimized state transfer\index{optimized state transfer}  chains from section
\ref{subsubsec:1.2.2} against this kind of ``manufacturing errors''.

(The work described in the remainder of this section of \cite{my64}
is based on the  
material published in \cite{my60,my62,my63}.)



 %
\section{Other theoretical approaches}
\label{sec:3}

In this section we review some studies dealing with the robustness of
quantum information transfer schemes. We are restricting ourselves to
the transfer of single-qubit states and we consider only static
randomness inherent in the system, that is, fabrication
errors. Furthermore, although interesting schemes involving different
kinds of qubit networks\index{qubit network} \cite{JS08,TPVx08,DFZ09,KT11,Tso11,AC12c}
have been suggested, we will discuss only strictly one-dimensional
systems. Despite these restrictions we are sure to have missed some
important contributions in this rapidly developing field.

(In \cite{my64} we also discuss the
contributions \cite{NPL04b,NPL04a,YBB10,Tsomokos2007,BFK12}.)

An influential early  paper is the study \cite{CRM+05} by De Chiara et
al who considered the influence of disorder on the state transfer
properties of the PST chain with linear energy 
spectrum\index{linear energy spectrum} \cite{CDEL04}.
 Both the nearest-neighbor couplings $J_i$ and the
local $z$ fields $h_i$ ($h_i=0$ in the ideal case) were assumed
random. Borrowing from the language of single-particle transport in a
one-dimensional tight-binding chain, one might call the disorder in
$h_i$ diagonal and the disorder in $J_i$ off-diagonal. These two kinds
of disorder are known to have fundamentally different effects on the
localization and transport properties of nearest-neighbor coupled tight-binding 
chains \cite{KM93,ITA94}. In fact, these two kinds of
disorder also turn out to be very different for the spin chains
 in \cite{CRM+05}: Coupling
constant disorder is the more detrimental the longer the chains
become, while magnetic field disorder apparently averages out for
longer chains. Numerical evidence and perturbation calculations for
weak disorder show that the fidelity \footnote{To be precise: the
  fidelity between the first and last spins, averaged over the Bloch
  sphere with respect to the first spin, calculated at $t_{PST}$ and
  averaged over the disorder.} is a decreasing function of the two
variables $N \varepsilon_J^2$ and $\varepsilon_h^2/N$, where
$\varepsilon_J$ measures the strength of the (relative) coupling
disorder and $\varepsilon_h$ does
the same for the (absolute) field disorder. Note that the ideal
coupling values $J_i=J \sqrt{i(N-i)}$ scale with the chain length so
that a given relative disorder strength $\varepsilon_J$ entails larger
absolute changes in the couplings for longer chains. In contrast the
random fields $h_i$ do not scale with $N$, and $\varepsilon_h^2/N$ is
what the central limit theorem yields for the variance of the (zero)
average field $\frac 1N \sum_i h_i$.

Burrell and Osborne \cite{BO07} investigated correlations in an infinite XX chain 
with random nearest-neighbor interactions and a random magnetic field
and showed that all correlations are exponentially suppressed outside of
an effective ``light cone'' whose radius grows at most logarithmically
in time (and hence is no light cone at all). This means that
information transfer out of a region of given size will take
exponentially long times in the limit of an infinite system. 
This is Anderson localization\index{Anderson localization} at work: In
one dimension all states go localized at arbitrarily small diagonal
disorder \cite{KM93}, but things are different for off-diagonal
disorder \cite{ITA94} since in that case there is always a delocalized
state at the center of the band. Localization effects were also
discussed by Keating et al \cite{KLM+07}, unfortunately without clear
distinction between the two kinds of disorder.
However, localization can be overcome for finite systems.
 In fact, it has been
shown \cite{AL09} how to employ quantum error correction\index{quantum
error correction} techniques to
send a qubit with high fidelity using several imperfect spin chains in
parallel, over distances large compared to the individual chain's
localization length.
In temporally fluctuating fields things are different, see
\cite{BEO09} for a study of localization properties in that case.

An early example for robustness considerations is the paper by
Kay \cite{Kay06}, which contains  a section about manufacturing
errors. This paper considers couplings beyond nearest neighbors  in
XX chains and also discusses the influence of, for example, timing
errors when reading out the transmitted state. 

 Three different scenarios of
quantum information transfer along XX spin chains are covered in
\cite{PNL10}.
 The first scenario involves sequential SWAP operations\index{SWAP operation}
effected by switching in turn every single spin coupling for an
appropriate duration. The second scenario employs the natural dynamics
of the PST system with linear spectrum without any external driving, while the
third one achieves adiabatic state transfer\index{adiabatic state transfer} by slowly switching all
even and odd couplings appropriately. The transfer times for all these
schemes scale as $\tau \sim N/J_{max}$, where $J_{max}$ is the maximum
coupling available. All three scenarios are studied in the presence of
diagonal (magnetic field) and off-diagonal (exchange coupling) static
randomness. It turns out that the sequential SWAP scheme is most
susceptible to randomness, especially of the off-diagonal type. The
linear-spectrum PST system  is more robust than the sequential SWAP scheme,
but the adiabatic state transfer 
scheme is most noise-tolerant, at least for the system sizes $(N \leq
51)$ studied.

 Ronke et al. \cite{RSA11} performed a comprehensive study of
 robustness of state transfer in short PST chains ($N \leq 15$) with linear
 spectrum \cite{CDEL04}.
The built-in perturbations considered
 were randomness in the nearest-neighbor couplings, site-dependent
 random magnetic fields, interactions between travelling excitations
 and unwanted next-nearest neighbor spin couplings. In addition,
 also handling errors, such as readout timing errors were studied. It was
 found that next-nearest neighbor spin couplings had the strongest
 detrimental affair on the quality of state transfer. The general
 behavior of the transfer fidelity was found to be consistent with an
 exponential decay with chain length and a Gaussian dependence on
 the disorder strength.

Bruderer et al. \cite{BFR+12} suggested a smart hybrid approach
unifying advantages of the fully-engineered and boundary-controlled
state transfer schemes. Their idea amounts to optimizing the temporal
structure (commensurate spectrum leading to perfect periodicity) and
the spatial structure (boundary-localized states insensitive to
perturbations from the interior of the chain) at the same  time. 

(In \cite{my64} a more detailed discussion
follows. Also discussed is the work of \cite{OSFx12}  on short
isotropic Heisenberg (XXX) antiferromagnetic chains, and of
\cite{LASx13}, where boundary-localized
states were created by erecting high magnetic field barriers on sites
close to the ends of a homogeneous XX chain, and the implementation of
 the PST scheme suggested by Christandl et
al. \cite{CDEL04} in an array of 19 laterally coupled
parallel waveguides \cite{PKKx13}.)

\section{Experimental implementations}
\label{sec:4}

In this section we review some of the experimental implementations
of state transfer\index{experimental implementations of state transfer}
using protocols based on spin chain channels and show their present
limitations. Every proposal for physical qubits that allows to couple
them permanently can be used to develop spin chains. Therefore, in
solid-state systems, there are proposals to implement qubit chains
using superconducting nanocircuits, such as charge qubits \cite{RFB05,Strauch2008},
Josephson junctions \cite{Tsomokos2007,Majer2007,Sillanpaa2007} or
flux qubits \cite{Lyakhov2005,Lyakhov2006}. The advances in semiconductor
technology allow to couple quantum dots, so there are proposals using
chains of charged quantum dots \cite{NPL04b,NPL04a,Petrosyan2006}
or alternatively, excitons in quantum dots \cite{DAmico2006,Spiller2007}.
Spin chains can be also simulated in optical lattices \cite{Dorner2003,Duan2003,Hartmann2007,Lewenstein2007,Ospelkaus2008,Lanyon2011,Bellec2012}
or with nuclear spin systems in NMR \cite{madi_time-resolved_1997,cory_ensemble_1997,gershenfeld_bulk_1997}.
Nitrogen vacancy centers in diamond \cite{YJG+11,cappellaro_coherence_2009,Neumann2010,Yao2012,PLB+13}
constitute another promising solid-state system. However, only very
few of these systems have actually developed into experimental implementations
of quantum spin channels and in particular NMR was the pioneer setup
for testing these protocols.

The main limitation to make these quantum channels a reality is decoherence\index{decoherence},
which not only affects the survival time of the quantum information
\cite{Zurek2003}, but also affects the distance over which it can
be transmitted
\cite{CRM+05,BO07,KLM+07,AL09,pomeransky_quantum_2004,apollaro_quantum_2007,alvarez_nmr_2010,alvarez_decoherence_2010,alvarez_localization_2011}.
Perfect or high fidelity state transfer can be obtained by many of
the theoretical methods described so far. However, if the ideal control
Hamiltonian or system for the state stransfer is affected by decoherence,
the transfer fidelity can be remarkably reduced. Decoherence effects
can come from either time dependent perturbations or even static ones.
In order to show this, let us consider the simplest quantum channel
of two qubits, where a SWAP operation\index{SWAP operation} transfers
the state from one qubit to the other. The experimental implementation
of a SWAP operation was first addressed within the field of liquid
state NMR \cite{Nielsen1998,madi_one-_1998,linden_how_1999}. However,
pioneering solid state NMR experiments performed by M\"uller, Kumar,
Baumann, and Ernst \cite{muller_transient_1974} can now be identified
as a SWAP operation. Even in this simple 2-qubit channel the swapping
oscillation is damped by a decoherence rate that depends on the rate
of interaction with the environment. Even worse, if the interaction
rate with the environment becomes larger than the ideal swapping frequency
between the two spins, the swap is frozen, manifesting an overdamped
dynamics due to localization effects of the initial excitation \cite{alvarez_decoherence_2010,alvarez_environmentally_2006,alvarez_signatures_2007}.

Ideally quantum communication is expected to be performed by means
of pure-state transfers. Consequently most of the theoretical approaches
focused on pure-state communication processes, but experimental realization
of pure states is a major challenge for present technologies. Just
recently a lot of progress is being made with superconducting devices,
semiconductor technologies, optical lattices and Nitrogen vancancy
centers where pure states can be generated and controlled, even at
the single qubit level \cite{Ladd2010,Cirac2012,Bloch2012,Blatt2012,Aspuru-Guzik2012,Houck2012}.
However, the first implementations of quantum computation were based
on ensemble quantum computing\index{ensemble quantum computing} using
mixed states that mimic pure-state quantum evolutions, in particular
with NMR setups \cite{cory_ensemble_1997,gershenfeld_bulk_1997,knill_power_1998}.
State transfer in a solid state system has been observed in a ring
of spins with dipolar (many-body) interactions\index{dipolar interactions}
\cite{pastawski_quantum_1995,pastawski_quantum_1996}. An initial
polarization localized in a specific spin of the ring propagates around
the ring and after a time related to the ring length, a constructive
interference reappears in the form of an echo \cite{pastawski_quantum_1995,pastawski_quantum_1996}.
This mesoscopic echo\index{mesoscopic echo} \cite{prigodin_mesoscopic_1994}
reflects the quantum nature of the finite quantum spin-ring. However,
the many-body nature of these spin-spin interactions made these systems
to sensitive to perturbations which strongly
reduce the amplitude of the echo \cite{Levstein1998}. In order to
improve the state transfer fidelity an effective XX interaction\index{XX model} Hamiltonian
\footnote{Many experimental references employ the term XY or planar interaction
for what we call XX interaction here.%
} has been experimentally implemented in a spin chain by
Madi et al \cite{madi_time-resolved_1997} by using global-pulse rotations
of the spins in a liquid state sample. In Ref. \cite{madi_time-resolved_1997},
the evolution of the initial excitation was monitored in all the spins
of the quantum channel. In that work two important features deserve
special attention. On the one hand it is relevant to be able to generate,
even if artificially, the simple one-body XX Hamiltonian for state
transfer because it increases the transfer fidelity in comparison
to more complicated many-body Hamiltonians. On the other hand, by
comparing the experimentally observed evolution of polarization transfer
to the ideal design, one can assess the decoherence effects reducing
the transfer fidelity. These decoherence effects originate from the
finite precision on control pulses and from interactions with external
degrees of freedom changing the coupling strengths between the spins
and/or their Zeeman energies.

Due to the non scalability of liquid state NMR, more recently, implementations
of spin chains were attempted in solid-state NMR \cite{Cap14,Cappellaro2007a,Cappellaro2007,Rufeil-Fiori2009,zhang_nmr_2009,cappellaro_coherent-state_2011,ramanathan_experimental_2011-1}
which mimic the XX state transfer evolution. This approach is based
on experimentally generating a double quantum Hamiltonian\index{double quantum Hamiltonian}
where the evolution of a locally prepared initial state can be mapped
to the one generated by an XX interaction \cite{Doronin2005}. However,
these systems do not allow individual addressing of the qubits, engineering
of the coupling strengths or local manipulations for generating PST.

Up to now, PST protocols could only be implemented in systems of very
few spins, mainly in trivial cases of 2 \cite{Nielsen1998,madi_one-_1998,linden_how_1999}
and 3 \cite{zhang_simulation_2005,Ajoy2012} spins interacting by
a homogeneous XX interaction. Again, the XX interaction was not natural
and in particular in these cases it was engineered by controlling
the spins individually and generating the desired effective Hamiltonian.
Similarly, but only requiring global control-pulses on a 6-spin system,
effective chains of 2, 3, and 6 spins were generated for implementing
PST protocols \cite{alvarez_perfect_2010}. In this case the selective
spin-spin coupling networks were created by exploiting selective quantum
interferences in the time domain to filter out the undesired couplings
while leaving intact the desired ones \cite{alvarez_perfect_2010}.
The process can be interpreted as a time-domain analog of Bragg gratings\index{Bragg gratings}
that filter the non-selected coupling strengths \cite{Ajoy2011,AC13}.
Alternative state transfer protocols with spin chains for achieving
arbitrarily high fidelities were constructed applying iterative state
transfer\index{iterative state transfer} \cite{BGB07} along chains
of 3 and 4 spins by controlling only the boundary spins \cite{zhang_iterative_2007,Zhang2009}.
Again, in all these cases decoherence\index{decoherence} by external
degrees of freedom or finite precision control changes the coupling
strengths or induces energy fluctuations of the spins affecting the
transfer fidelity.

The decoherence\index{decoherence} effects also are the main limitation
on the chain lengths. Decoherence effects increase as the number of
qubits increases \cite{alvarez_nmr_2010,alvarez_localization_2011,krojanski_scaling_2004,krojanski_reduced_2006,krojanski_decoherence_2006,lovric_decoherence_2007}.
The sensitivity of the quantum states grows with the number of spins, causing
imperfections, disorder or
external influence on the  couplings within the spin channel to induce
localization\index{localization of the quantum information} of the quantum information \cite{CRM+05,BO07,KLM+07,AL09,pomeransky_quantum_2004,apollaro_quantum_2007,alvarez_nmr_2010,alvarez_localization_2011,burrell_information_2009}.
These localization effects were recently observed experimentally in
three-dimensional spin-network topologies with about 7000 spins, where
the localization effects were induced by finite precision control
of the quantum gates driving the information transfer \cite{alvarez_nmr_2010,alvarez_localization_2011}.
Thus it is clear that the only way of building quantum computers or
quantum simulators has to be based on developing robust methods of
controlling the information and in particular the state transfer \cite{Hauke2012}.

\section*{Acknowledgements}
A. Z. and O. O. acknowledge support from SECYT-UNC and CONICET. A. Z.
thanks for support by DAAD, G. A. A. for support by Alexander von Humboldt
Foundation; both acknowledge the hospitality and support of Fakult\"at Physik
of TU  Dortmund.

\bibliographystyle{spphys}
\bibliography{/home/stolze/BIBLIO/jsbas_def,/home/stolze/BIBLIO/general,/home/stolze/BIBLIO/mypapers,/home/stolze/BIBLIO/qcomp,/home/stolze/BIBLIO/random_chain,Bib-book-chapter-2013-Stolze}

\end{document}